# Peculiar solar sources and geospace disturbances on 20–26 August 2018

**A. A. Abunin[1,2], M.A. Abunina[1], A. V. Belov[1], I. M. Chertok[1]**

**Abstract** On the approach to minimum of Solar Cycle 24, on 26 August 2018, an unexpectedly strong geomagnetic storm (GMS) suddenly occurred. Its $D_{st}$ index reached −174 nT, that is the third of the most intense storms during the cycle. The analysis showed that it was initiated by a two-step long filament eruption, which occurred on 20 August in the central sector of the solar disk. The eruptions were accompanied by two large-scale divergent ribbons and dimmings of a considerable size and were followed by relatively weak but evident Earth-directed coronal mass ejections. In the inner corona, their estimated speed was very low of about 200–360 km s$^{-1}$. The respective interplanetary transients apparently propagated between two high-speed solar wind streams originated from a two-component coronal hole and therefore their expansion was limited. The resulting ejecta arrived at the Earth only on 25 August and brought an unexpectedly strong field of $B_t \approx 18.2$ nT with a predominantly negative $B_z$ component of almost the same strength. The geospace storm also manifested itself in the form of a peculiar Forbush decrease (FD). Its magnitude was about 1.5 %, which is rather small for the observed G3-class GMS. The main unusual feature of the event is that large positive bursts with an enhancement up to 3% above the pre-event level were recorded on the FD background. We argue that these bursts were mainly caused by an unusually large and changeable cosmic ray anisotropy combined with lowering of the geomagnetic cutoff rigidity in the perturbed Earth's magnetosphere under conditions of the cycle minimum.

## 1. Introduction

Nonrecurrent geomagnetic storms (GMSs) and Forbush decreases (FDs) of galactic cosmic rays are driven by two types of solar activity phenomena caused by a sudden explosion of non-potential magnetic field configurations: (1) magnetic eruptions in active regions (ARs) accompanied often by noticeable flares and (2) eruptive filaments outside ARs (see for a review Belov, 2009; Webb and Howard, 2012; Kumar and Badruddin, 2014; Vial and Engvold, 2015; Gopalswamy *et al.*, 2015). The common properties of these two types of eruptions are that AR eruptions also include eruption of filaments. Solar filaments are dense and cool plasma held by magnetic fields over the polarity inversion line in the chromosphere and lower corona. Therefore, in the Hα and extreme ultraviolet (EUV) images they look like extended dark formations. The non-AR filament eruptions (so-called quiescent prominences) also contain some low-intense flare features, in particular such as dimmings and long-existing post-eruptive

-------------------------------------------------------------------

Corresponding author: Artem Abunin (abunin@izmiran.ru)

[1]Pushkov Institute of Terrestrial Magnetism, Ionosphere and Radio Wave Propagation (IZMIRAN), Troitsk, Moscow 108840, Russia
[2]Kalmyk State University, Elista 358000, Russia

ORCID
A.A. Abunin      https://orcid.org/0000-0003-1427-3045
M.A. Abunina     https://orcid.org/0000-0002-2152-6541
A.V. Belov       https://orcid.org/0000-0002-1834-3285
I.M. Chertok     https://orcid.org/0000-0002-6013-5922

arcades with diverging ribbons. In eruptions of both types, the direct agents carrying the magnetized plasma from the Sun to the Earth are coronal mass ejections (CMEs) sometimes of the halo-type and their interplanetary counterparts (ICMEs and magnetic clouds). It is clear that the AR eruptions occur in stronger magnetic fields and are therefore more energetic. Respective ICMEs arrive to the Earth faster and cause more intense GMSs and FDs than non-AR filament eruptions (Chertok *et al.*, 2013). The occurrence rate of both types of eruptions and associated non-recurrent space weather disturbances is highest during the solar cycle maximum. However, such events may also occur during the declining phase of a solar cycle. They are rather rare but nevertheless of significance. At the decline phase of Solar Cycle 24, which is the weakest cycle of the century, at least two outstanding ensembles of solar and geospace events were observed.

One of the most powerful outbursts of solar flare activity, which happened on 4–10 September 2017, attracted the attention of solar and solar-terrestrial researchers on various aspects and many related articles have already been published (*e.g.*, Yang *et al.*, 2017; Chertok *et al.*, 2018; Cohen and Mewaldt, 2018; Gopalswamy *et al.*, 2018; Hou *et al.*, 2018; Verma, 2018, and references therein). The outbursts occurred due to the extremely rapid development and increasing complexity of the magnetic structure of AR12673 during its passage over the western half of the visible disk. Within one week, the Sun produced numerous C-class flares, 27 M-class flares, and four major X-class flares, including the two strongest flares in Cycle 24: SOL2017-09-06T12:02 (X9.3-class) and SOL2017-09-10T15:58 (X8.2-class). Three halo-type CMEs were observed over the course of this activity. The effects on the geospace environment were a number of solar energetic particle (SEP) events including one ground-level enhancement (GLE) on 10 September, as well as moderate to strong GMSs and FDs.

Another remarkable event happened a year later at the approach to the very deep minimum of Cycle 24. On 26 August 2018, under rather quiet background conditions, a strong GMS with the third by the minimum value of the $D_{st}$ index in the cycle of –174 nT (http://wdc.kugi.kyoto-u.ac.jp/dstdir/index.html; see also Watari, 2017) and a rather weak (≈ 1.5%) for this GMS but highly peculiar FD suddenly occurred. Such a strong G3 GMS was not foreseen by forecasters. According to the NOAA Space Weather Prediction Center (SWPC) (https://www.swpc.noaa.gov/news/g1-minor-geomagnetic-storm-warning-effect; see also ftp://ftp.swpc.noaa.gov/pub/warehouse/2018/WeeklyPDF/prf2243.pdf), a minor G1 GMS was only expected. Recall that according to NOAA Space Weather Scale (https://www.swpc.noaa.gov/noaa-scales-explanation#), the minor G1 and strong G3 classes of GMSs correspond to indexes $K_p = 5$ and 7, respectively. The analysis showed, the indicated space weather disturbances were caused by a non-AR central filament two-step eruption on 20 August which was accompanied by two consecutive Earth-directed low-speed CMEs. The arrival of the respective ICMEs to the Earth in 4–5 days resulted in a fairly strong magnetic field with a predominant southern $B_z$ component.

The galactic cosmic rays (CRs) interact with ICMEs approaching the Earth and their variations, often starting before the beginning of GMSs, depend on the structure, orientation, magnetic topology of transients and generally serve as precursors of GMSs. These variations include pre-decrease and pre-increase of the CR intensity, as well as of the CR anisotropy changing in magnitude and direction. In addition, FD itself often begins before a GMS, as was the case in the August 2018 events (see, *e.g.*, Belov, 2009; Cane, 2000; Dorman, 2009 for a review).

It is known that the time profile of FD, occurring in a combination with GMS, is often distorted due to variations of the CR anisotropy and lowering of the geomagnetic cutoff rigidity. In particular, these effects can lead to a short enhancement or burst of the CR count rate on the background of the main FD observed at some neutron monitors (NMs). Several such events recorded during Cycle 24 are analyzed in the recent papers by Evenson *et al.* (2017), Gil *et al.*

(2018), Munakata *et al.* (2018), Samara *et al.* (2018). Even those CR bursts that stand out well at the FD recovery phase do not usually reach the pre-event count rate level. The CR enhancement took place also during the August 2018 FD and revealed some interesting peculiarities. Firstly, at several mid-latitude NMs, the enhancement significantly exceeded the pre-FD level and reached a positive magnitude of 3%. Secondly, at the Moscow (IZMIRAN) NM, the enhancement consisted of two large positive bursts separated by a time interval of several hours. We will show that these peculiarities were mainly caused by the unusually large CR anisotropy (up to 3%) and strong variations of its direction, although variations of the CR cutoff rigidity during GMS also contributed to the observed effects. The fact that this event occurred on the approach to minimum of the solar cycle under high intensity and relatively low modulation of CRs also played a role.

This paper addresses the 20–26 August 2018 events and is organized as follows. In Section 2, the solar source of the geospace disturbances is analyzed including the two-step non-AR filament eruptions and associated CMEs. Section 3 considers the peculiar features of FD and resulting GMS. In particular, appearance of significant bursts of the CR count rate on the FD background is considered. Then in Section 4, a summary and discussion are presented.

## 2. Two-step solar eruption

For most of 2017 and especially 2018, the Sun was spotless and without significant filaments. Recurrent coronal holes (CHs), including the near-equatorial and trans-equatorial ones, and related activity prevailed. On 20 August 2018, the Sun as a whole had the appearance typical for the approaching minimum of Cycle 24, but contained some distinctive features. As the 193 Å images of the EUV telescope *Atmospheric Imaging Assembly* (AIA; Lemen *et al.*, 2012) on board the *Solar Dynamic Observatory* (SDO; Pesnell *et al.*, 2012) show (Figure. 1a), two very small active regions AR12718 and AR12719 were located south of the helioequator. However, these two ARs were of no interest since they did not have any activity manifestations during that day. The main attention should have been attracted to a relatively small non-AR dark filament (quiescent prominence) and its southward continuation in the form of a channel along the polarity inversion line of the large-scale background magnetic field. The filament and channel were situated in the northern half of the disk, near the central meridian, between the northeastern CH and southwestern CH881. As can be seen from Figure 1, the latter consisted of two parts and generally was quite extensive in heliolongitude, the leading part being located somewhat west and the structured trailing component east of the eruption zone. The cold matter of the northern filament itself was clearly visible in the Hα line (Figure 2), and the filament channel apparently contained heated plasma. The total length of these structures was quite large and exceeded 0.8 Rs.

Solar and solar terrestrial observers and forecasters have not paid enough attention to either the filament or the channel. Meanwhile, a flux rope corresponding to these structures underwent a sequential two-step eruption on 20 August, that had significant space weather consequences. As the SDO/AIA movies presented at https://sdo.gsfc.nasa.gov/data/dailymov.php and the 193 Å images shown in Figure 1b demonstrate, the eruption above the channel occurred at first at about 08–09 UT. This eruption 1 was accompanied by the formation of long-lived coronal structures characteristic of such phenomena. Two large-scale luminous flare-like ribbons appeared on both sides of the pre-existing channel and two dimmings (areas of reduced brightness) D1 and D2 originated near ends of the ribbons. The ribbons are treated as sets of legs of the loops that compose the post-eruption arcade almost not visible in this case. The dimmings are the result of the plasma outflow from footpoints of the erupted flux rope located above the channel. Both ribbons and dimmings were long-lived and existed for many hours, changing their configuration.

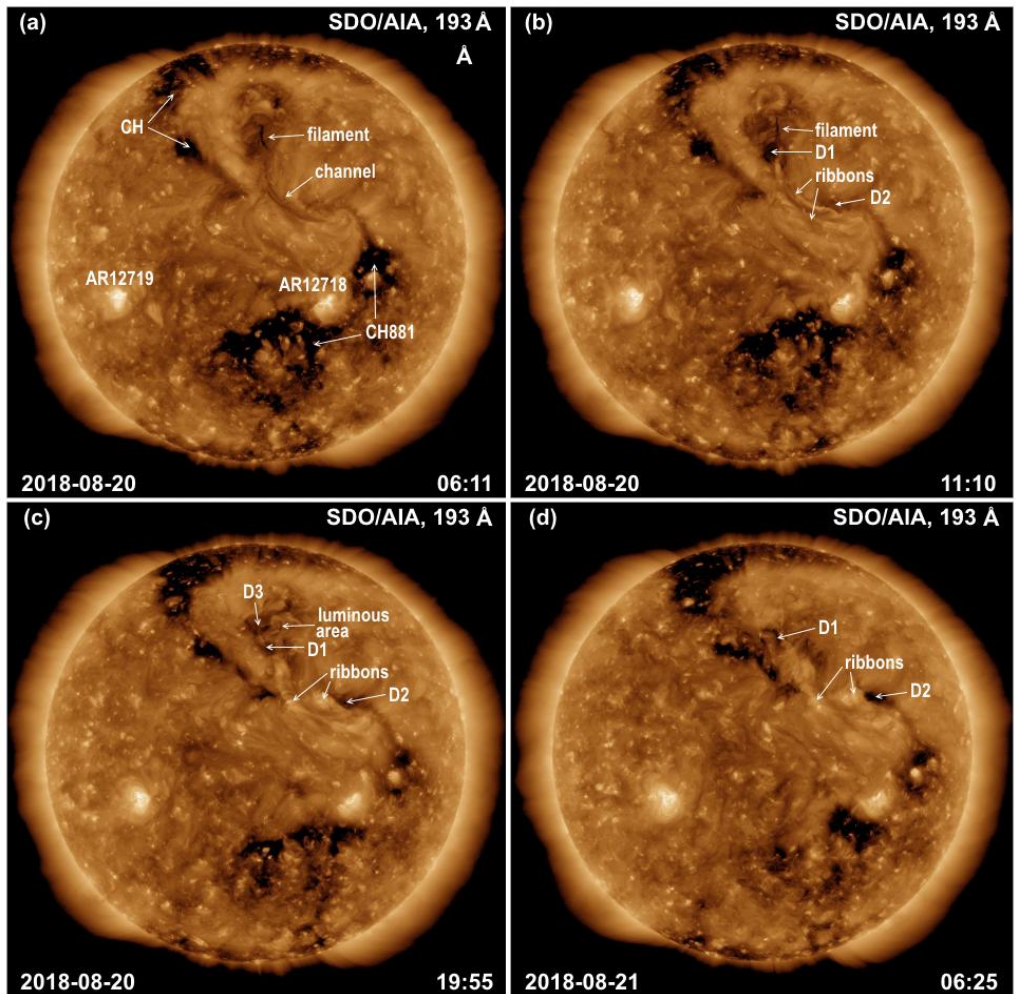

**Figure 1** The SDO/AIA 193 Å images before eruptions (**a**), after eruption 1 (**b**), shortly after (**c**) and 12 hours after (**d**) eruption 2. Relevant coronal holes (CHs), eruptive filament, filament channel, flare-like diverging ribbons, and core dimmings (D1, D2, D3) are indicated by *arrows*.

The ribbons were diverging, i.e. the distance between them increased with time, that corresponds to the formation of increasingly higher post-eruptive loops.

It is essential that eruption 1 from the filament channel and the adjacent northern dimming D1 almost did not affect the northern filament itself (Figure 1b). However, the Hα movie of the Big Bear Solar Observatory (BBSO, ftp://ftp.bbso.njit.edu/pub/archive/2018/08/20/) (see also Figure 2) shows that at 17–18 UT, the northern part of the Hα filament has disappeared. This means that an eruption 2 occurred. As the SDO/AIA 193 Å movie and images (Figure 1b and 1c) show, this eruption was accompanied by the appearance of a luminous area in the filament place and by the formation of an additional northern dimming D3. This indicates a significant restructuring of the magnetic field in this region of the corona, characteristic of eruptions. It cannot be ruled out that some structural elements related to the southern filament channel were also involved in this relatively weak eruption 2, although the SDO/AIA movies show no obvious signs of this. The diffuse diverging ribbons and dimming D1 and D2 continued to be observed for many hours after eruption 2 (Figure 1d).

It is noteworthy that both eruptions 1 and 2 did not cause any meaningful increases of the total solar soft X-ray flux recorded with the *Geostationary Operational Environmental Satellites* (GOES). The 1–8 Å emission remained within the background level of A2–A3 (ftp://ftp.swpc.noaa.gov/pub/warehouse/2018/2018_plots/xray/). According to the SWPC report, any manifestations in the radio emission were also not fixed on 20 August. This means that the flare processes were expressed very weakly in these events.

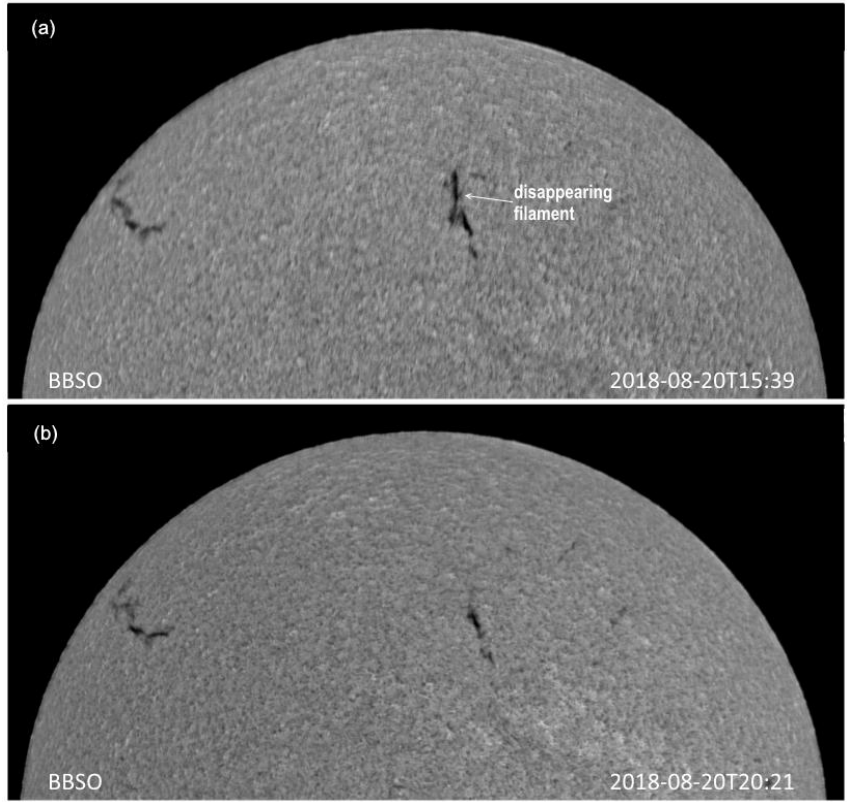

**Figure 2** The BBSO images of the Hα erupting (disappearing) filament before (**a**) and after (**b**) eruption 2.

Nevertheless, there are clear observational evidence that both eruptions 1 and 2 were accompanied by Earth-directed CMEs, although of rather small-sized and low-speed. We used information from the Solar Eruptive Event Detection System (SEEDS; http://spaceweather.gmu.edu/seeds/) providing, in particular, the automatically processed white-light images and movies recorded with the *Large-Angle Spectrometric Coronagraph C2* (LASCO/C2; Brueckner *et al.*, 1995) on the *Solar and Heliospheric Observatory* (SOHO; Domingo *et al.*, 1995) and with the *Outer Coronagraph COR2* (Howard *et al.* 2008) aboard the *Solar Terrestrial Relations Observatory Ahead* (STEREO-A; Kaiser *et al.* 2008). Recall that SOHO is permanently in the Lagrange point L1 and STEREO-A was on 20 August 2018 in the vicinity of the Earth's orbit at an angle of 108° east of the Sun-Earth line. The location of SOHO was not favorable and the position of STEREO-A, on the contrary, was suitable for observations of CMEs originating near the central meridian of the solar disk visible from the Earth. The situation is shown in Figure 3, where the representative running-difference LASCO/C2 (left row) and STEREO/COR2 (right row) images are displayed for both eruptions. On the LASCO image, a CME from the eruption 1 (Figure 3a) manifested itself very weakly: only faint amorphous glow is visible above the near-equatorial sectors of the occulting disk of the coronagraph. At the same time, the STEREO-A difference image (Figure 3b) reveals a more or less distinct loop-like CME1 with a central position angle (PA) of about 285°. It seems that , for technical reasons, only the northern half of this CME is visible. The eruption 2 was accompanied by a somewhat more significant CME2 (Figure 3c, 3d). The SOHO/LASCO difference image displays it as a noticeable halo luminous around the occulting disk. Accordingly, in the STEREO-A/COR2 images, the eruption 2 produce a developing structured CME2 rising approximately in the same direction as CME1, i.e. at PA ≈ 285–290°. Since the centre of the erupted Hα filament was placed at the northern heliolatitude of about 40°, the indicated PA mean either that CME2 was deflected from the radial direction toward the ecliptic plane by some coronal magnetic structures, for example, associated with the northeastern CH (see Manchester *et al.*, 2017), or that a flux rope located over the filament channel was involved in eruption 2 together with the Hα filament. The angular size of CME1 and CME2 in the sky plane of STEREO-A/COR2 was about 20–30°.

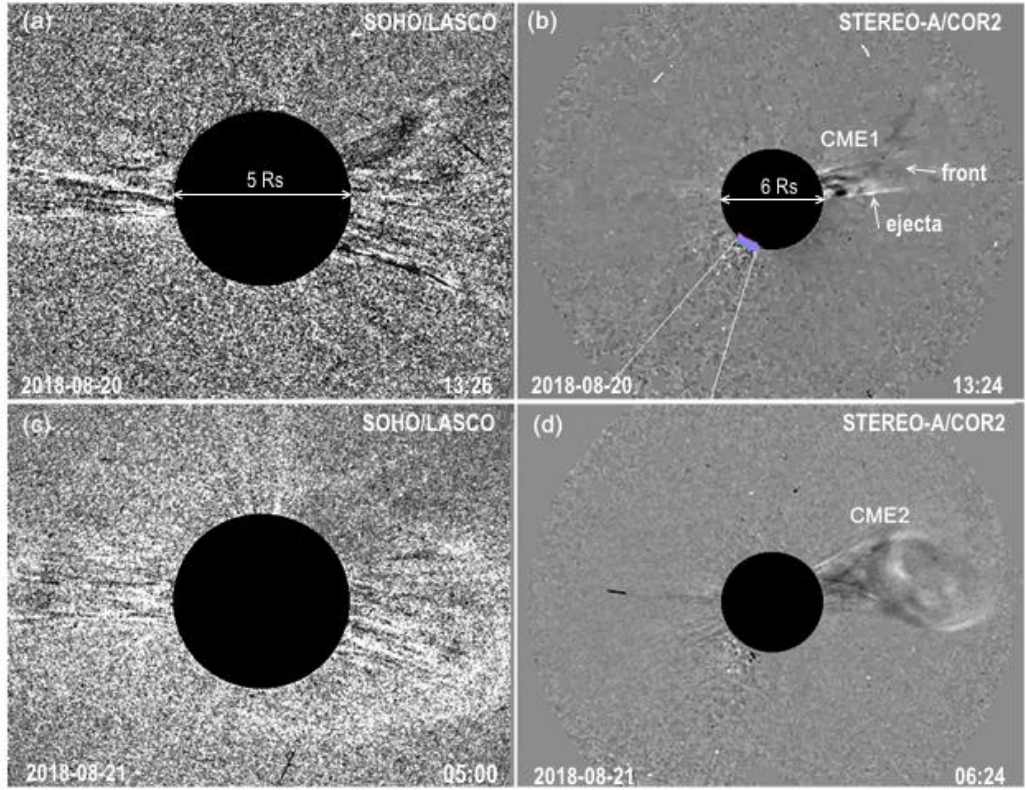

**Figure 3** The CME1 and CME2 on the SEEDS running difference images of the SOHO/LASCO/C2 (*left column*) and STEREO-A/COR2 (*right column*) coronagraphs. In (**a**, **b**), the diameter of the occulting disk of both coronagraphs is indicated in solar radii (Rs) for scaling. In (**b**), an outlined southeast sector with the blue arch at the bottom is not related to CME1.

The restricted but not too poor visibility of the two CMEs described above allows us to make approximate height-time measurements of their frontal structures in the sky plane of STEREO-A/COR2. The results are presented in Figure 4. For CME1, it was possible to determine several positions of the front and ejecta marked in Figure 3b. Their trajectories are consistent with the time of eruption 1 around 08–09 UT and reveal a low speed of 240–340 km s$^{-1}$. More detailed information is possible to obtain for CME2. These data indicate the time of eruption 2 of about 16–18 UT, close to the disappearance time of the Hα filament. The speed of CME2 is also quite small, but variable. During the first 7–8 hours it was ≈110 km s$^{-1}$ and then it increased up to ≈225 km s$^{-1}$. Taking into account the STEREO-A location at an angle 108° to the Sun–Earth line, it is possible to convert these speeds into the space speed in the Earth direction. As a result the CME1 and CME2 speeds are of order 255–360 km s$^{-1}$ and from ≈115 km s$^{-1}$ to ≈235 km s$^{-1}$, respectively. These low estimated coronal speeds mean that both CME1 and CME2 and their interplanetary counterparts ICMEs should be transported to the Earth by the background solar wind.

**3. Geospace disturbances**

Location of the sources of both eruptions 1 and 2 near the central meridian, the character of visibility of the respective CMEs from the SOHO/LASCO and STEREO-A/COR2 coronagraphs, their observed PAs and the width of the order of several tens of degrees in the sky-plane give grounds to assume that both CME1 and CME2 could be directed toward the Earth. The journey of low-speed CMEs/ICMEs, which were carried by the background solar wind, to the Earth's orbit lasted about 4–5 days, as it was expected.

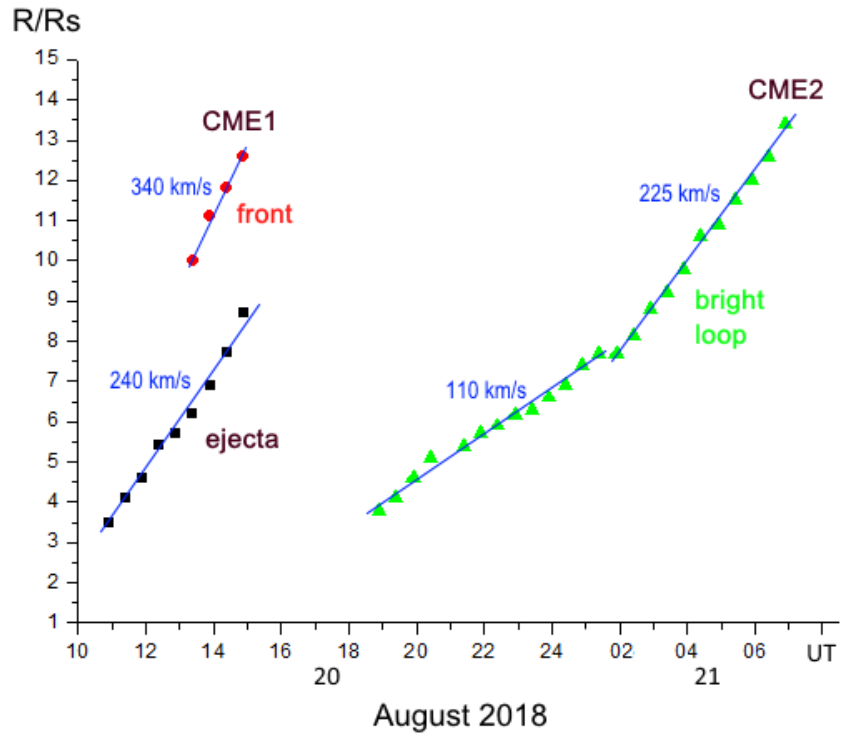

**Figure 4** The height-time trajectories and speeds of the CME1 and CME2 frontal structures in the STEREO-A/COR2 sky plane.

In situ solar wind measurements by the *Deep Space Climate Observatory* (DSCOVR, https://www.ngdc.noaa.gov/dscovr/portal/index.html#/vis/summary), which is located at the Lagrangian L1 point near the Earth, yield to the following plausible picture (Figure 5). Several particular regions can be distinguished. By 24 August, the Earth exited from the high-speed solar wind stream associated with the leading part of CH881. ICME1 from eruption 1 (region I) arrived at the Earth on 24 August at about 06 UT and manifested itself by a slight but sharp increase of the interplanetary magnetic field (IMF) up to 7 nT. ICME2 from eruption 2 propagated through the wake of ICME1. The turbulent region II indicates the interaction of the frontal structure and sheath of ICME2 with the tail part of ICME1. The hit of the main body of ICME2 (region III) on the Earth's magnetosphere at noon on 25 August was apparently much more direct and powerful than in the ICME1 case. As a result, ICME2 brought to the Earth a fairly strong IMF (up to $B_t \approx 18$ nT) with a southern (negative) $B_z$ component of almost the same strength. These magnetic field parameters reached their peaks at around 02 UT on 26 August and remained at these extreme levels for 11 and 8 hours, respectively. Such a prolonged negative $B_z$ component indicates that the ICME2 flux rope near the Earth was oriented almost perpendicular to the ecliptic plane in accordance with the solar filament orientation. The ICME2 body presents certain signatures of a magnetic cloud, in particular, the IMF enhancement with a smooth rotation of the field direction and variable temperature. The rise of speed and temperature in regions IV and V can be interpreted accordingly as areas of interaction a high-speed stream from the trailing part of CH881 with the tail of ICME2 and then as the CH high-speed stream itself. The presence of two increases of density and pressure, peaked in regions III and IV at 04–06 and 13 UT on 26 August, reflects a complex structure of the incident wind caused by the large heliolongitudinal extension and the intermittent configuration of the trailing part of CH881 (see Figure 1). In general, this plausible picture is consistent with the WSA-ENLIL+Cone model run for the 20–27 August 2018 period that can be seen at

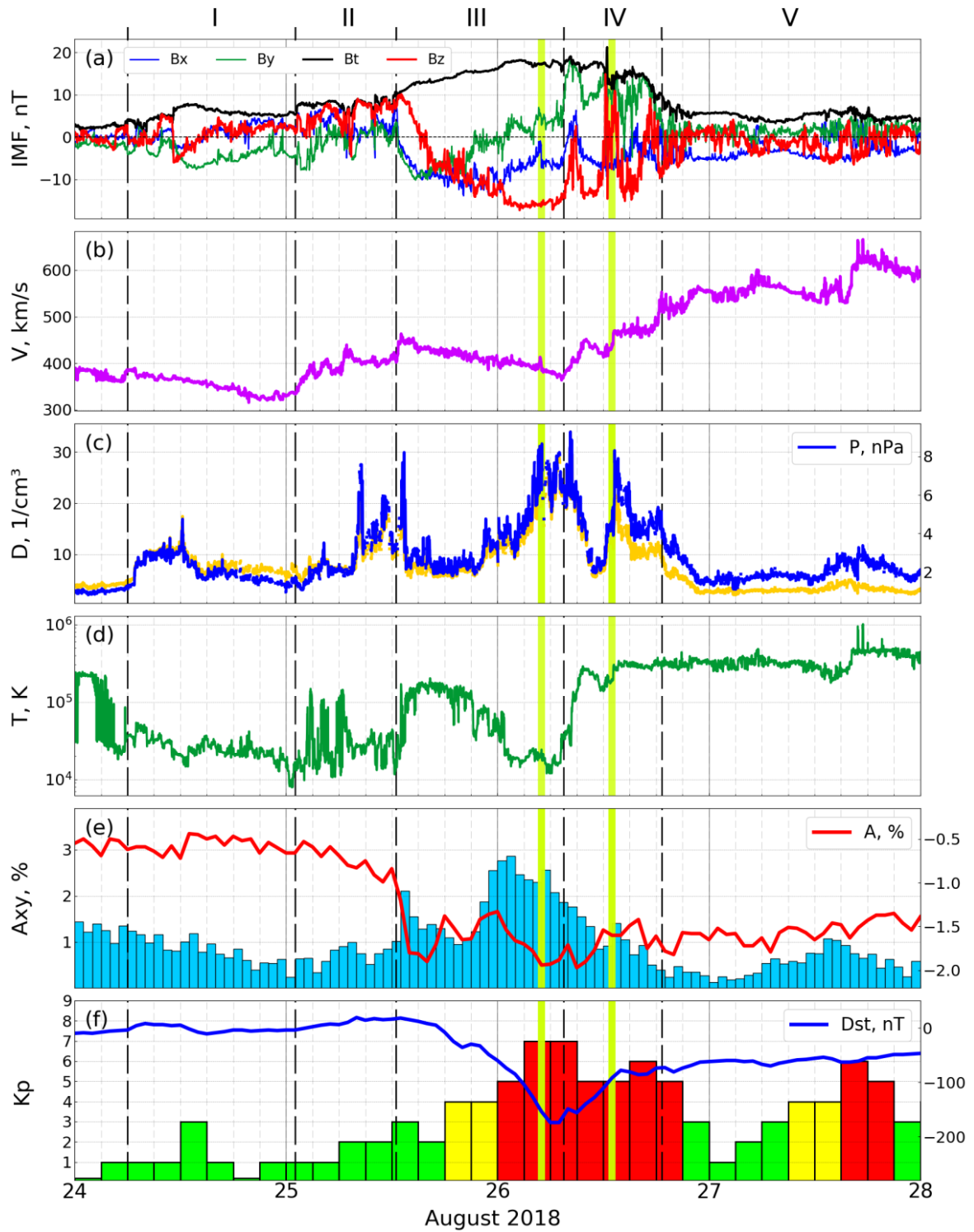

**Figure 5** *In situ* near-Earth DSCOVR data on the magnetic field (**a**), solar wind speed (**b**), density/pressure (**c**), temperature (**d**) for the 2018 August 24–27 period. Time profiles of the 10 GV cosmic ray density $A$, the value of the equatorial anisotropy $A_{xy}$ (**e**) and the geomagnetic $D_{st}$, $K_p$ indexes (**f**). In this and subsequent figures, the *vertical dashed lines* outline the regions of the main near-Earth disturbances. The *yellow vertical lines* correspond to peaks of two positive bursts of the CR count rate at the MOSC monitor (see below).

https://iswa.ccmc.gsfc.nasa.gov/IswaSystemWebApp/StreamByDataIdServlet?allDataId=145432 1065.

Now, we outline data and approaches that are used by us to analyze the Forbush-decrease (FD) and geomagnetic storm (GMS). The FD magnitude is taken from the IZMIRAN Database of the Forbush-Effects and Interplanetary Disturbances (Database FEID, http://spaceweather.izmiran.ru/eng/dbs.html; Abunin *et al.*, 2019). The maximum flux depression corresponds to a cosmic ray (CR) rigidity of 10 GV and is determined by data from the world network of neutron monitors using the global survey method (GSM) (see Belov *et al.* (2018) and references therein). The GSM allows to combine data of the entire network of neutron monitors (Neutron Monitor Data Base (NMDB); www.nmdb.eu) and get the main characteristics of the variations in the near-Earth space outside of the atmosphere and magnetosphere. The meaning of this procedure is that it gives the density of the CR variations and the 3D vector of the CR anisotropy, as they would be measured by a hypothetical detector outside of the atmosphere and the magnetosphere of the Earth capable of detecting the 10 GV particles coming from all directions. The used version of GSM is based on that the CR variations at any observation point may be presented as the sum of the zero-order and the first spherical harmonic of an expansion in Legendre polynomial. Isotropic part of the CR intensity (or density) is a zero harmonic in spherical expansion. Equatorial and north-south anisotropy are the components of the first spherical harmonic. Two indexes $D_{st}$ and $K_p$ characterize GMS. The hourly values of the storm time disturbance index $D_{st}$, calculated from the data of four low-latitude geomagnetic observatories and characterizing the effect of the global equatorial ring current, are presented at http://wdc.kugi.kyoto-u.ac.jp/dstdir/index.html. The planetary 3-hr $K_p$ index is calculated using data from geomagnetic stations located at moderately high geomagnetic latitudes mainly in the northern hemisphere. The $K_p$ data are downloaded from ftp://ftp.gfz-potsdam.de/pub/home/obs/kp-ap/wdc/.

ICME1 did not cause noticeable space weather disturbances, whereas ICME2 was accompanied by a peculiar FD and a strong GMS (Figure 5e, 5f). The FD began to develop as a gradual decrease of the CR density starting at 04 UT on 25 August and coincided with the increase of the solar wind speed, density, and IMF in region II which indicates the interaction of the frontal structure of ICME2 with the tail part of ICME1. A sharp decrease of the CR density occurred at 13 UT on 25 August, immediately after entering region III and the associated further increase of IMF and the solar wind density in the ICME2 body. The magnitude of FD ≈ 1.5 % was observed after 4 hours. Then the CR density displayed noticeable variations, firstly in the form of a two-hump increase to 1.4%, and then for two days including regions IV and V, in the form of a gradual weakening. In general, the evolution of FD reflected the passage of ICME2 near the Earth, and the sharp and especially gradual decrease of the CR density occurred many hours before the onset and peak of GMS.

Despite the low speed (of about 400 km s$^{-1}$), but due to a relatively strong total IMF $B_t$ and a significant negative $B_z$ component, not typical for an approach to minimum of the solar cycle, the main body of CME2 (region III) caused a considerable GMS (Figure 5f), which started at 13 UT on 25 August and reached a level of $K_p = 5$ at 00 UT on 26 August. The later onset of GMS in regard to the onset of FD was due to the fact that the $B_z$ component reached sufficiently large negative values only several hours after the onset of the ICME2 body (region III). The GMS magnitude peaked at 07–08 UT on 26 August with $D_{st} ≈ -174$ nT , $K_p ≈ 7+$, i.e. this GMS became the third by intensity in Cycle 24, at least with regard to the $D_{st}$ index. At the beginning of the next day, the GMS intensity dropped to $D_{st} ≈ -50$ nT and remained near this disturbed level for about two more days, i.e. in the regions IV and V where the fast solar wind from the trailing part of CH881 contributed. The $K_p$ index increased during 27 August and at 15–18 UT reached another peak $K_p ≈ 6$, while the $D_{st}$ index remained almost unchanged.

Shown in Figure 6 are time profiles of the relative CR count rate (intensity) registered by four NMs with different locations and cutoff rigidities: the Russian Moscow NM (MOSC; coordinates 55.47° N, 37.32° E; geomagnetic cutoff rigidity $Rc ≈ 2.43$ GV), the Antarctic Jang Bogo NM (JBGO; 74.6° S, 164.2° E; $Rc ≈ 0.3$ GV); the North American Nain NM (NAIN;

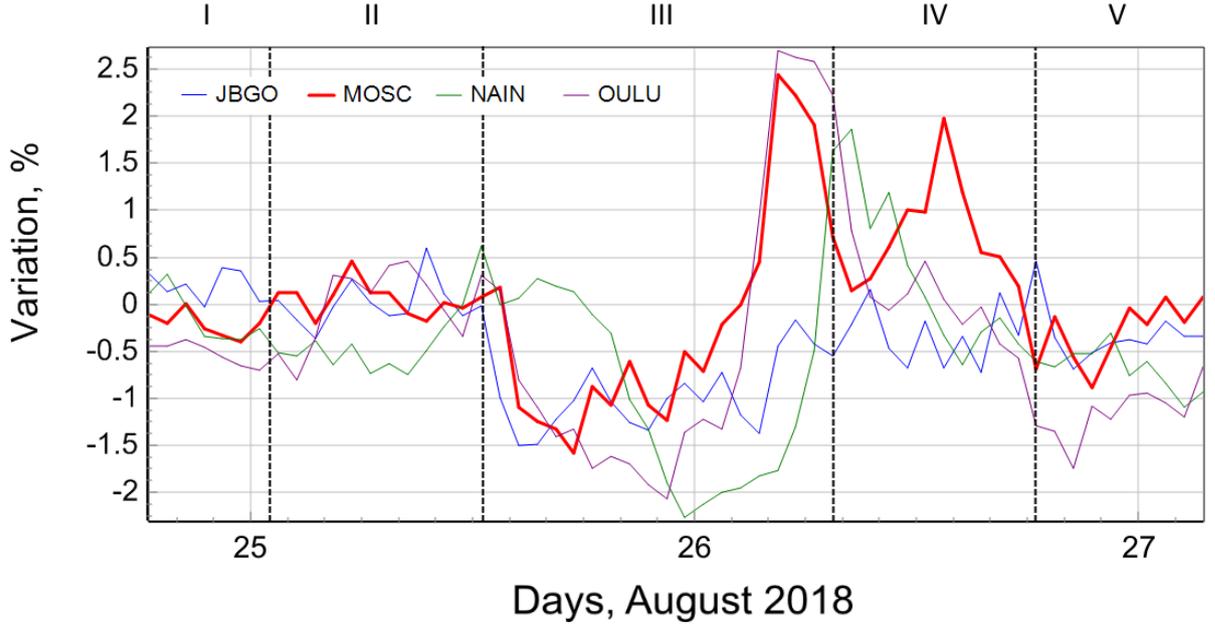

**Figure 6** Different manifestations of the Forbush decrease: a simple FD at JBGO and peculiar FDs at the MOSC, NAIN, OULU neutron monitors with the positive count rate enhancements. The *vertical dashed lines* outline the regions I–V of the main near-Earth disturbances dedicated in Figure 5.

56.55° N, 61.68° W; $Rc \approx 0.3$ GV); the Finish Oulu NM (OULU; 65.05° N, 25.47° E; $Rc \approx 0.8$ GV). One can see that these profiles are very different. At the JBGO NM (blue profile), there is only a decrease of the CR density, characteristic of a simple FD. The remaining three mid-latitude NMs demonstrate a very unusual FD. The profile of the NAIN NM (green line) displays a delayed decrease followed by a single significant positive burst of the CR intensity up to 1.8% at 07–11 UT on 26 August. The largest burst up to 3% above the pre-FD count rate level was recorded at the OULU NM (purple line) at 04 UT. In time, it was slightly ahead of the GMS peak moment as determined by the $D_{st}$ index at 07–08 UT, both coinciding with the tail section of the ICME2 body (region III). Below we will show that a similar single large positive enhancements were registered approximately at the same time at a number of the mid-latitude NMs. Even more peculiar count rate profile was fixed by the Moscow NM (red curve). Here, after a short decrease down to $\approx -1.5\%$, two positive bursts of CR intensity up to 2.4 and 2 % lasted 3.5 and 2 hrs and separated by an interval of about 9 hr took place. In this case, the maximum of the first burst coincided in time with the largest OULU enhancement. The second MOSC burst peaked at 14 UT on 26 August, during the interaction of ICME2 with a high-speed stream from the trailing part of CH881 (region IV). We will argue that the unusual large positive enhancements indicated above are due to the combined action of the CR anisotropy and the magnetospheric variations.

We used the global survey method to complement the CR density time profile with a vector anisotropy diagram depicted in Figure 7 (see Belov *et al.*, 2018). Here, for each hour, the magnitude and direction of the equatorial component of the CR anisotropy $A_{xy}$ are represented by the magnitude and direction of the blue vectors. The thin purple straight lines connect the same points in time on the two curves every 6 hours and show to which portion of the CR time profile (brown curve) the respective vectors belong. Variations of the north-south component of the CR anisotropy $A_z$ are shown by the vertical green arrows. The equatorial CR anisotropy for another FD is calculated and presented in the same way in the article of Samara *et al.* (2018). The CR anisotropy is sensitive to changes in the structure and parameters of the solar wind, in particular, such as ICMEs, shocks, high-speed streams from CHs and their interaction areas. This was observed in the August 2018 events. From the Figure 7 data it is clear that at the beginning of the event and during the principal FD, the main anisotropy vector was directed from south-south-east that is close to the galactic CR anisotropy direction, typical for quiet conditions in the Solar Cycle 24 (see Belov, 2009). The first short-term but noticeable change in the direction of the equatorial anisotropy took place at about 06 UT on 25 August in the turbulent magnetic field of region II during interaction of ICME2 with the tail part of ICME1 and associated gradual decrease of the CR density. A short peak of the equatorial anisotropy up to 2% (see also Figure 5e) occurred several hours later, at 14 UT and coincided with the sharp initial phase of FD and the entrance of the Earth into the ICME2 body (the beginning of region III). A more intense and prolonged rise of the equatorial anisotropy up to $A_{xy} \approx 2.9\%$ was observed on the morning of 26 August slightly before the first burst at the MOSC CR time profile and during the growing phase of GMS (the second half of region III). This value is 5–6 times higher than the typical $A_{xy}$ magnitude for quiet periods. At this stage, the north-south anisotropy $A_z$ became directing toward the south and gradually increased in value, reaching $\approx 2\%$ near the peak of the first MOSC burst and also significantly exceeding the normal magnitude for quiet periods. At this stage, the north-south anisotropy $A_z$ became directing toward the south and gradually increased in value, reaching $\approx 2\%$ near the peak of the first MOSC burst and also significantly exceeding the normal level. Furthermore, at the interval between two MOSC bursts and during approaching the second MOSC burst, the equatorial anisotropy direction changed most significantly. At this time, the interplanetary magnetic field was the most turbulent and changeable in region IV during interaction of ICME2 with a high-speed stream from the trailing part of CH881 (see Figure 5). In general, both the equatorial and north-south components of the CR anisotropy were abnormally large from 22 UT on 25 August to 11 UT on 26 August. The large $A_z$ anisotropy, directed from north to south, provided an additional contribution to the increase of the CR density inside FD for NMs located in the northern hemisphere. From the end of 26 August, when the Earth entered the high-speed stream from the trailing part of CH881 and perhaps from the northeastern CH (region V), no large changes of the CR density and anisotropy were observed.

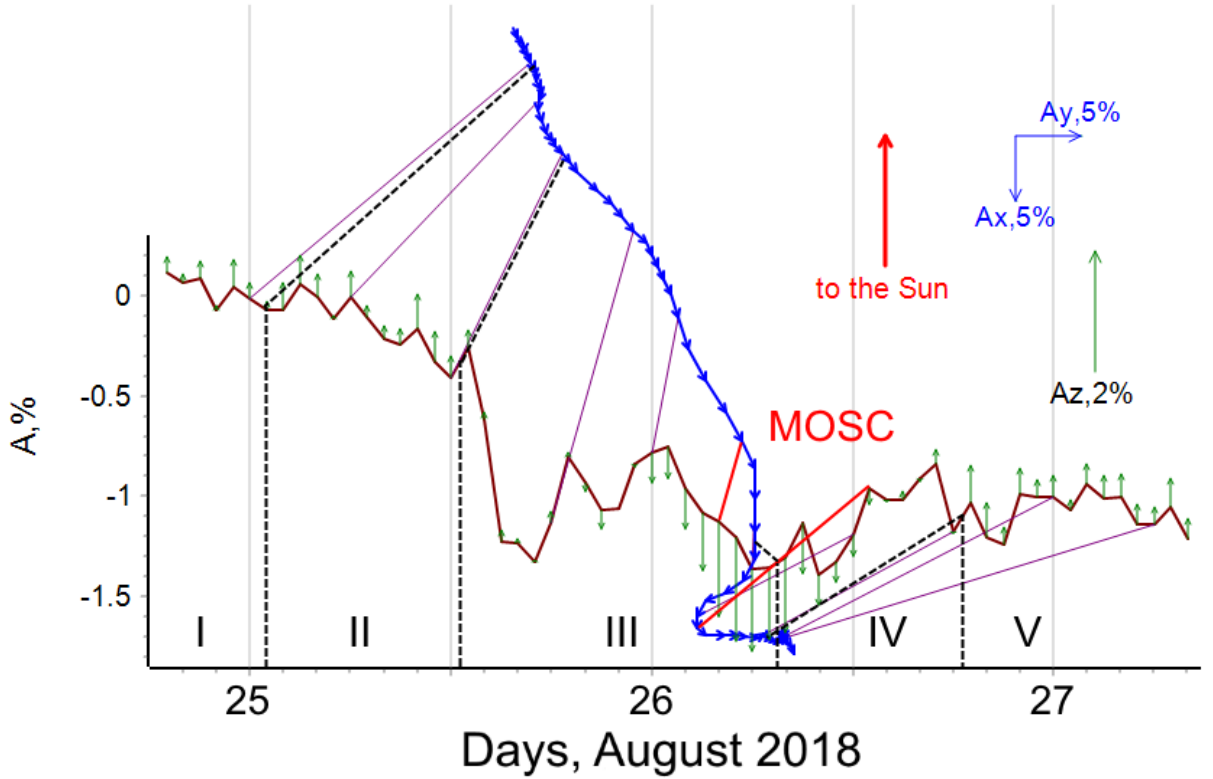

**Figure 7** Behavior of the 10 GV cosmic ray density $A$ (*brown curve*) and anisotropy. The *blue* linked vector diagram displays the magnitude and direction of the equatorial anisotropy $A_{xy}$ at each hour, the vector's direction corresponds to direction of CR flux; thin *violet straight lines* connect the same points in time on two curves every 6 hours; *vertical green arrows* show changes of the north-south anisotropy component $A_z$. The $A_{xy}$ and $A_z$ scales are indicated at the *right on the plot*. The *red lines* correspond to the first and second bursts on the MOSC neutron monitor. The *black dashed lines* outline the regions I–V of the main near-Earth disturbances dedicated in Figure 5.

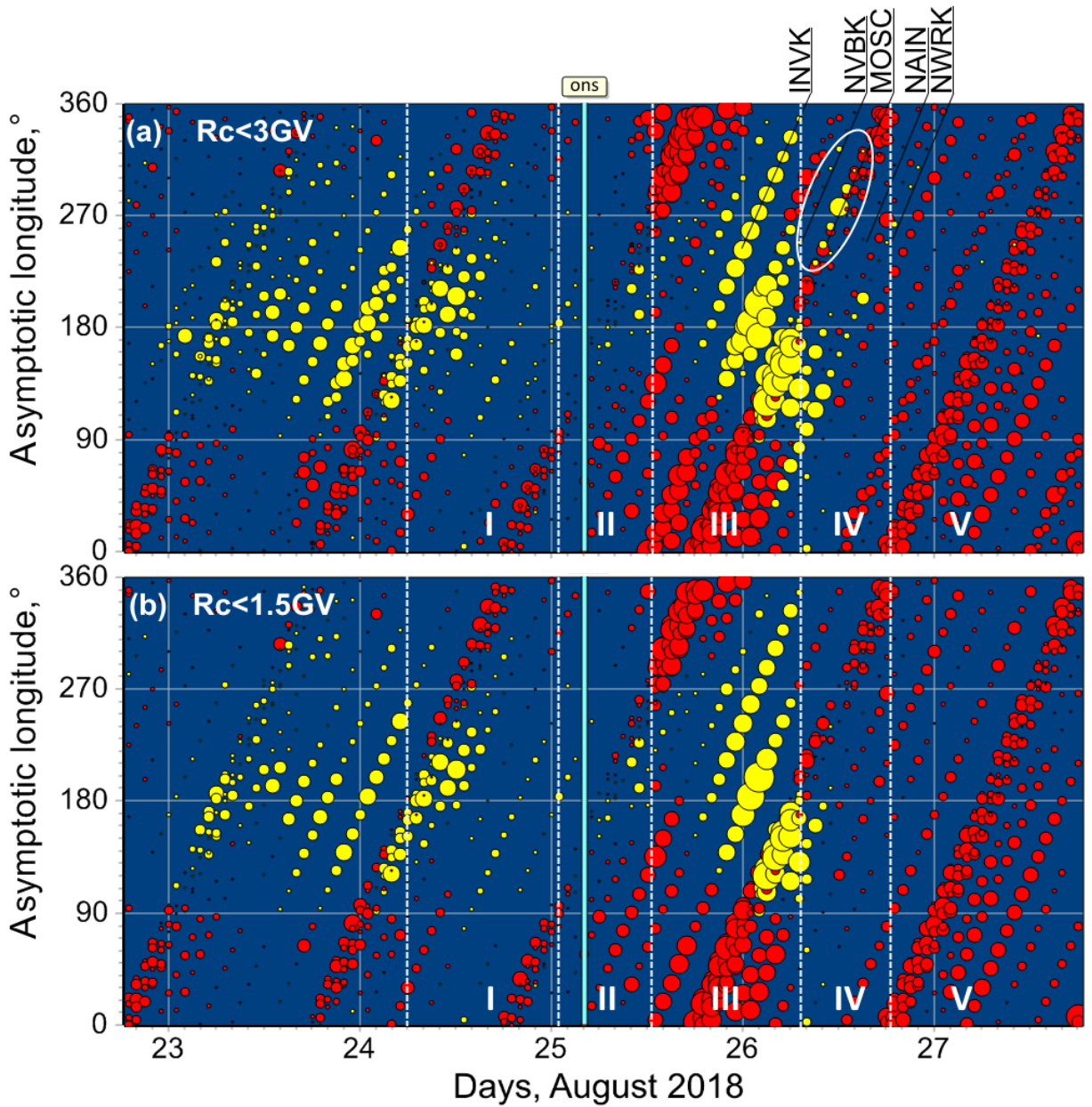

**Figure 8** Distribution of the cosmic ray variations in the asymptotic longitude for the 22–27 August 2018 period by data of 15 neutron monitors with the geomagnetic cutoff rigidity $Rc < 3$ GV (**a**) and 11 NMs with $Rc < 1.5$ GV (**b**). Decreases and increases of the CR intensity in relation to the quiet base period are shown by *red and yellow circles*, respectively. The size of the circles is proportional to the variation magnitude. Each *inclined* straight line refers to an individual neutron monitor and the line sequence is repeated every day. The vertical blue line marks the FD onset. In panel (**a**), the *white oval* delineates the second burst at the MOSC and NVBK NMs. The *vertical dashed lines* outline the regions I–V of the main near-Earth disturbances dedicated in Figure 5.

Additional information on variations is obtained using so called the ring station method (Papailiou *et al.*, 2012). It allowed us to consider how the data of a series of NMs are distributed on the "asymptotic longitudes – time" plane (Figure 8). The asymptotic longitude displays the direction of particle arrival recorded by a concrete detector in geocentric coordinates. It depends on the geographical coordinates of the station, the time of day and the effective drift of particles in the Earth's magnetic field. Decreases and increases of the CR intensity in relation to the quiet base period are shown by red and yellow circles, respectively. The size of the circles is proportional to the variation magnitude. Each inclined straight line refers to an individual NM and their sequence is repeated every day. For ordinary events (including those in which all NMs display the count rate enhancements below the pre-event level on the FD background), only red cycles should be present to the right of the vertical onset line in pictures similar to Figure 8. Note that similar representations of the cosmic ray variations are used by other authors and have been published, for example, in the papers of Munakata *et al.* (2005) and Kuwabara *et al.* (2006).

Data of two groups of NMs with different geomagnetic cutoff rigidity ($Rc$) are presented for the 23–27 August period in Figure 8. Panel (a) is formed by data of 15 NMs with $Rc < 3$ GV, including MOSC. For example, data of the Inuvik NM (INVK), Novosibirsk NM (NVBK), Nain NM (NAIN) and the Newark NM (NWRK) are marked. Prior to the FD onset, on 23–24 August, we see the usual longitudinal-time distribution, characteristic of the daily variation in a quiet period. The FD (large red cycles) begins on 25 August not simultaneously at different stations, and on the day side of the Earth there is no significant CR decrease at all. A remarkable enhancement (enlarged yellow cycles) is clearly visible at longitudes of 100°–250° in the period from 18 UT on 25 August to 10 UT on 26 August that corresponds to the first burst on the MOSC count rate time profile (see Figure 6). In addition to the MOSC NM, this burst is also displayed on the INVK, NAIN and NWRK data but at slightly different time that is due to the different geographical location of the stations. The second positive burst, which is delineated by the white oval, is also present around 12–14 UT by the MOSC and NWBK data at longitudes of about 280° and 320°, respectively. In this depiction, it looks weaker than the first burst which is represented by data of many NMs.

In this case, a large increase of the CR intensity is observed, in particular, at NMs with the highest cutoff rigidity, which are included in the ring station method, such as Moscow, Newark, Novosibirsk. The question that arises is how the anomalous distribution that is seen in Figure 8a will change in transition to NMs with fewer rigidity restrictions, i.e. to what extent the observed effect is a variation of the magnetospheric origin. In order to answer this question, we present a similar figure, but only for 11 high-latitude NMs with $Rc < 1.5$ GV (Figure 8b), for which the magnetospheric variations have almost no effect on the counting rate changes. One can see that the second burst disappeared because the MOSC and NVBK data are not presented here due to the accepted rigidity limitation. At the same time, the clear manifestations of the first burst on the FD background at longitudes of 100°–200° remained although they somewhat weakened. This weakening and narrowing of the region of positive variations mean that the CR cutoff rigidity variations were involved in the formation of the first burst, although they played a minor role. Thus, the magnetospheric variations are not the main cause of the first positive burst, although they are also involved in it. The CR anisotropy makes a main contribution to the first positive burst, and the relative contribution of the magnetospheric variations to the second burst is greater than to the first one. The CR anisotropy probably also played a role in the origin of the second positive burst, since this burst occurred during the most significant direction changes of the CR equatorial anisotropy (see Figure 7).

## 4. Summary and Discussion

The presented data reveal that the two-step non-AR filament eruptions observed on 20 August 2018 are obviously a solar source of the significant geospace disturbances that occurred in 5−6 days. This is evidenced by the absence of any other eruptive events on the Sun in the right time and the characteristic features of eruptions 1 and 2. The filament and its channel were located near the central meridian and their eruptions (disappearance) were accompanied by such inherent signatures as the long diverging luminous ribbons and adjacent dimmings. It is especially important that two CMEs were recorded after these eruptions. Analysis of the data of two coronagraphs on board SOHO and STEREO-A from different viewpoints showed that these CMEs could be directed to the Earth. Within the 15-solar-radius part of the STEREO-A/COR2 coronagraph fields of view, their PA width was of the order of several tens of degrees, and their speed was quite small, not exceeding 360 km s$^{-1}$. In addition, modest visibility of CMEs on both coronagraphs indicates a low plasma density inside. With such parameters throughout the entire route from the Sun to the Earth, CMEs/ICMEs should be carried by the background solar wind between relatively fast winds from the leading and trailing parts of CH881 and, most importantly, experience only weak lateral expansion.

It is known that the expansion rate of CMEs/ICMEs largely determines the strength of the magnetic field that they transport to the Earth (*e.g.*, Manchester *et al.*, 2017). For example, Grechnev *et al.* (2014) argued that a very strong magnetic field (up to 56 nT) in the magnetic cloud that caused the superstorm on 20 November 2003, was due to an unusually weak expansion of the disconnected spheromak in an enhanced-density environment constituted by the tails of the preceding ICMEs. According to Gopalswamy *et al.* (2014) and Manoharan *et al.* (2018), the weak Solar Cycle 24 is characterized by a reduced total pressure in the heliosphere that usually leads to the relatively large expansion of ICMEs. However, the degree of expansion is positively correlated with the radial propagation speed of CMEs/ICMEs (Gopalswamy *et al.*, 2014). Along with cases of continuous expansion of ICMEs, there are events in which the expansion of transients was limited, due to which their magnetic field could remain intact over long distances (*e.g.*, Manoharan, 2010). In our case of 20–26 August 2018, the circumstances were unusual and the CME speed was so low that the opposite effect probably could took place. At the entire Sun-Earth path, both relatively weak transients propagated inside and between fast wind streams from leading and trailing parts of CH881 that inhibited large ICME expansion. In addition to this, when propagating to the Earth, ICME1 and ICME2 apparently interacted with each other, as well as with the wind streams from CH881. Such interactions could lead to compression, i.e. to amplification of the internal magnetic fields (see Manchester *et al.*, 2017). Thus, the weak expansion of CMEs / ICMEs in the corona and solar wind and their interactions perhaps resulted in a relatively strong magnetic field in the resulting transient formed near the Earth. Two additional factors appeared to be decisive for the occurrence of the observed significant geospace disturbances, in particular, of GMS: (a) a perfect hit of the transient on the Earth's magnetosphere and (b) the southward orientation (negative $B_z$ component) of the total magnetic field. In our case, all these factors are apparently jointly realized. Without this, the emergence of such a strong storm as G3 would be impossible.

The GMS of August 2018 with $D_{st} \approx -174$ nT, $K_p \approx 7+$ is unusually strong for an approach to minimum of a solar cycle. Moreover, it was accompanied by a remarkable FD with the magnitude of about −1.5% that is relatively small for such a significant GMS. One of the possible factors explaining such a combination of GMS and FD is as follows. As is known (*e.g.*, Cane, 2000; Belov, 2009), the intensity of GMSs is determined by the local parameters of the interplanetary transients directly in the region of their interaction with the magnetosphere, while the FD magnitude depends on the global characteristics of the transients approaching the Earth. In this case, the relatively weak FD probably occurred because ICMEs had small sizes (about 20–30°). This conjecture is consistent with the assumption that the strong field in ICME2 was due to relatively weak expansion under its transport from the Sun to the Earth.

The important feature is that the anomalously large equatorial and north-south CR anisotropy of about $A_{xy} \approx 2.8\%$ and $A_z \approx 2.35\%$ was observed during this weak FD. In these respects, the August 2018 event is rare and even unique for solar cycle minimums. According to the IZMIRAN Database FEID covering the 1957–2018 period (http://spaceweather.izmiran.ru/eng/dbs.html; Belov *et al.*, 2017), only a few GMSs of the same or greater intensity took place at the minima of cycles 19–23. Further, about 5,000 FDs with the magnitude < 1.5% occurred during these 61 years, but among them, there were only 5 events with such an anomalously large equatorial and north-south CR anisotropy, as in the August 2018 event. For FDs with the size ≈ 1.5%, the average equatorial anisotropy is $A_{xy}$ ≈ 1.3%. One more distinctive feature of this rather weak FD is a very sharp drop of the CR density to its minimum which occurred at 13–15 UT on 26 August during the entry into the ICME2 body (region III). This may explain the unusually large CR anisotropy. It is reasonable to assume that the latter is also due to a large and prolonged increase of the $B_z$ magnetic component.

The small decrease of the CR density during FD, combined with a large increase of anisotropy, created unusual profiles of the count rate variations at several mid-latitude NMs with a large positive enhancement up to ≈ 3% above the pre-event level. Most often, large GMSs are observed during high solar activity in a combination with sufficiently deep FDs which are also accompanied by the CR enhancements. However, the magnitude of the enhancements in such events usually does not exceed the pre-FD level (Dorman, 2009). Our analysis showed that the August 2018 positive enhancements was caused by the combined action of the variations of the CR anisotropy and lowering of the CR cutoff rigidity but the main contribution to the positive bursts was given by the anomalously large CR anisotropy and its variations. This happened during the interaction of the Earth's magnetosphere firstly with the body of ICME2 (region III) and then during its contact with the high-speed solar wind stream from the trailing part of CH881 (region IV). It should be kept in mind that the mid-latitude NMs have a rather high sensitivity both to changes of the CR anisotropy and to variations of the geomagnetic cutoff rigidity, although the magnetospheric variations are usually rather small even during large GMSs. In addition, during very low solar activity, the CR modulation is minimal and the intensity of low-energy CRs is greater than usually (see Dorman, 2009).

The unique features of the August 2018 event, discussed above, created conditions for another remarkable peculiarity of FD. At the Moscow NM, the large CR enhancement over the FD background consisted of two intense positive bursts up to 2.4 and 2% of which only the first one coincided in time with the enhancement at other mid-latitude NMs and the second one took place in several hours. This peculiarity is most likely due to combination of the following additional factors. Firstly, during the main geospace disturbances, the Moscow NM turned out to be in a favorable longitudinal zone. Secondly, the maximum increase of the CR anisotropy close to the peak moment of GMS and the largest changes of the equatorial anisotropy by its direction were shifted in time. Thirdly, in this case, the spatial displacement of asymptotic longitudes of the Moscow NM due to rotation of the Earth fortunately coincided with the temporal direction changes of the CR anisotropy vector. This may explain the presence of the second MOSC burst, which almost did not appear on other NMs.

Regarding to the magnetospheric variations of CRs, conditioned by a decrease of the cutoff rigidity, it should be kept in mind that usually at such mid-latitude NMs as MOSC they are rather small even during strong GMSs. Most often, large GMSs are observed during high solar activity when the CR modulation is strong. The 2018 August GMS occurred during a period of very low activity, when the modulation of the CR was minimal and the intensity of low-energy CRs (in the NM sense) was greater than usual. During such periods, the CR magnetospheric variations are much larger than during periods of high solar activity. The amplitude of the magnetospheric variations is related to the geomagnetic cutoff rigidity and the response function, which shows the relative contribution of CRs of various energies in the intensity observed by NMs (see Dorman, 2009). In the period of low solar activity and minimal CR modulation, the response functions in the range of small rigidities increase significantly. These factors explain the large CR variations of the magnetospheric origin in the event under consideration.

Variations of the CR anisotropy in value and direction reflect in more detail the structure of interplanetary disturbances than CR density variations. In particular, the interplanetary shocks, manifested as sudden GMS commencements (SSC) , and boundaries of magnetic clouds are normally clearly seen in behavior of anisotropy (see Belov, 2009). In the August 2018 events, there was no shock (SSC), and the first short-lived increase of the equatorial CR anisotropy and the sign change of the north-south anisotropy at the middle of 25 August coincided with the sharp onset of the main FD caused by interaction of the Earth's magnetosphere with the ICME2 body (region III in Figure 5). The maximum equatorial and north-south CR anisotropy and the associated first positive burst of the count rate at a number of the middle-latitude NMs, including MOSC, as well as the GMS peak, took place at the beginning of 26 August during the strongest total magnetic field $B_t$ and $B_z$ component in ICME2 as it approached the area of interaction with a high-speed stream from the trailing part of CH881 (boundary between regions III and IV). The second positive burst at the Moscow NM occurred on 26 August at about 13 UT simultaneously with the most dramatic direction changing of the equatorial CR anisotropy straight in the region IV incidentally coinciding with the space displacement of asymptotic longitudes of the Moscow NM due to rotation of the Earth. Time coincidence of two positive bursts at the MOSC monitor with two enhancements of density and pressure (see Figure 5) are explained by that the fast wind from the extended trailing part of CH881 had a complex structure and its crowding onto the tail of ICME2 led firstly to rise of the spatial gradient of the CR density, that caused increase of the equatorial and north-south anisotropy, and then to sharp changes in the direction of anisotropy.

Of course, the main reason of the enhanced anisotropy is the large values of the spatial gradient of the CR density. This conjecture is consistent with a rather large value of the maximum hourly decrease of the CR density of about 0.6 % / hr during the sharp onset of the main FD at 12–14 UT on 25 August. This indicates a significant radial gradient associated with the entry into the ICME2 body. However the enlarged CR anisotropy continued for a long time that can't be associated with a radial gradient only. Apparently, the contribution of other components of the CR gradient was also significant and manifested longer. It can be assumed that the Earth moved for a quite long time in a region of a large latitudinal gradient near the southern edge of ICME2. Recall that the central positional angle of ICME2 was located northward of the solar equator.

Thus, the event under consideration has become outstanding due to the following features. (1) A large geomagnetic disturbance was accompanied by FD with a modest decrease of the CR density. (2) The large CR anisotropy during this FD, although it was not a record one, caused positive enhancements significantly exceeding the decrease of the CR count rate. (3) A strong GMS occurred at the background of a high, weakly modulated CR intensity, that led to unusually large CR variations of magnetospheric origin at mid-latitude stations.

The remarkable solar eruptions and geospace disturbances of 20–26 August 2018 (as well as the outstanding events associated with a surprising large burst of the flare activity on 4–10 September 2017 mentioned in Introduction) expand our ideas of what powerful and peculiar events may occur in the minimum of the solar cycle. Study of the August 2018 event shows that, on the approach to a cycle minimum, the space weather forecasters should pay attention to even minor non-AR filament eruptions and associated small and slow CMEs. The respective ICMEs, with a successful hit on the Earth's magnetosphere and the presence of a negative $B_z$ component, can cause noticeable geospace disturbances, including substantial GMSs and peculiar FDs. It is important to monitor FDs and the CR anisotropy because they contain leading information about ICMEs approaching the Earth and potential GMSs.

An article by Chen *et al.* (2019) has just been published which also addresses the 20–26 August 2018 solar-terrestrial events. This and our papers are mutually complementary. They consider slightly various aspects of these events but threat the quiet-region filament and associated slow CME/ICME as a cause of the unexpected strong GMS. Both papers argue (albeit somewhat differently) that a sufficiently strong magnetic field in the CME flux rope at 1 AU (in ICME2 according to our terminology), necessary for occurrence of GMS, was due to the limited expansion of the ejecta by its interaction with fast solar winds originating from adjacent CHs. Unlike Chen *et al.* (2019), we additionally pay great attention to variations of galactic cosmic rays and analyze a peculiar FD with strong enhancements above the pre-event level on its background initiated by an unusually large and changeable cosmic ray anisotropy combined with lowering of the geomagnetic cutoff rigidity in the perturbed Earth's magnetosphere. In general, two articles together give a more complete picture of the 20–26 August 2018 events occurred at the minimum stage of Cycle 24.

In another recently published article, Mishra and Srivastava (2019) analyze in detail the development and linkage of CMEs of 20 August 2018, classifying them as stealth. However, they themselves show that their source was a coronal plasma channel and an overlying flux rope. Our analysis reveals that this was accompanied by such characteristic low-coronal signatures of eruptions as large-scale divergent ribbons and dimmings. Therefore, it isn't reasonable to consider the analyzed transients as stealth.

**Acknowledgments** We are grateful to an anonymous reviewer for useful remarks and comments. The authors thank the BBSO, DSCOVR, GOES, SDO/AIA, SEED, SOHO/LASCO, STEREO-A/COR2, SWPC, NMDB, and other related teams for the open data used in this study. All data sources used in producing the results presented in this article are quoted in Sections 2 and 3. This research was partially supported by the Russian Foundation of Basic Research under grants 17-02-00308 and 17-02-00508 , by the Russian Science Foundation under grant 15-12-20001, and by the Complex Program 19–270 of the Russian Ministry of Education and Science.

**Disclosure of Potential Conflicts of Interest** The authors declare that there is no conflict of interest for publishing this research results in Solar Physics.